\documentclass[preprint,showpacs,preprintnumbers,amsmath,amssymb,superscriptaddress]{revtex4}

\usepackage{graphicx,amsfonts}
\usepackage{epsfig}
\usepackage{dcolumn}
\usepackage{bm}
\usepackage{color}
\definecolor{red}{rgb}{1,0.0,0}
\begin{document}

\title{
Explicit parity violation in $SU(2)_L\otimes SU(2)_R\otimes U(1)_{B-L}$ models
}


\author{ Henry Diaz}%
\email{hdiaz@uni.edu.pe}
\affiliation{Facultad de Ciencias, Universidad Nacional de Ingenier\'\i a (UNI), av. Tupac Amaru s/n, Lima - Rimac 15333, Per\'u}

\author{ E. Castillo-Ruiz}%
\email{ocastillo@uni.edu.pe}
\affiliation{Facultad de Ciencias, Universidad Nacional de Ingenier\'\i a (UNI), av. Tupac Amaru s/n, Lima - Rimac 15333, Per\'u}

\author{O. Pereyra Ravinez }%
\email{opereyra@uni.edu.pe}
\affiliation{Facultad de Ciencias, Universidad Nacional de Ingenier\'\i a (UNI), av. Tupac Amaru s/n, Lima - Rimac 15333, Per\'u}

\author{V. Pleitez}%
\email{v.pleitez@unesp.br}
\affiliation{
Instituto  de F\'\i sica Te\'orica--Universidade Estadual Paulista \\
R. Dr. Bento Teobaldo Ferraz 271, Barra Funda\\ S\~ao Paulo - SP, 01140-070,
Brazil
}

\date{28/07/2021}
\begin{abstract}
Here we propose models with $SU(2)_L\otimes SU(2)_R\otimes U(1)_{B-L}$ electroweak gauge symmetry in which parity (or charge conjugation) is broken explicitly and the interactions of left- and right-handed fermions are completely different from each other at any energy scale. In these sort of models there is no restoration of parity at high energies. 
However, as in all left-right symmetric models, the electroweak charged interactions of quarks and leptons appear to be only left-handed at low energies, because the right-handed interactions are suppressed by the mass of the respective charged vector bosons and/or because these interactions may be much weaker than the left-handed ones.  
\end{abstract}


\maketitle

\section{Introduction}
\label{sec:intro}

The basic assumption of the left-right symmetric models is that the electroweak gauge symmetry is
$G_{LR}\equiv SU(2)_L\otimes SU(2)_R\otimes U(1)_{B-L}\otimes \mathcal{P}(\mathcal{C})$, i.e., the lagrangian is also invariant under a generalized parity $\mathcal{P}$~\cite{Pati:1974yy,Mohapatra:1974hk,Mohapatra:1974gc,Senjanovic:1975rk,Senjanovic:1978ev, 
Minkowski:1977sc,Davidson:1978pm,Marshak:1979fm} (or/and a generalized charge conjugation $\mathcal{C}$~\cite{Maiezza:2010ic}) and at high energies, these generalized parities  are restored. As a consequence, right-handed currents and  right-handed neutrinos must exist in nature. However, the former must be suppressed and the latter have to be massive particles. 
In early works on left-right symmetric models, the parity was broken softly~\cite{Mohapatra:1974gc}, but it could also be done  spontaneously~\cite{Senjanovic:1975rk}. This sort of models are well motivated because parity (and charge conjugation) violation is accommodated but not explained in the context of the standard model (SM). 

Besides, another possibility which we will explore here, is that no generalized parity symmetry exist at all: as in the SM, the parity (or charge conjugation) is explicitly broken at any energy. This case is less elegant than the case in which the breakdown of these symmetries is spontaneous, but it cannot be excluded, at least from the phenomenological point of view.
We recall that the only known source  of $C\!P$ (Charge Conjugation and Parity) violation is the phase in the CKM (Cabibbo-Kobayashi-Maskawa) matrix i.e., it is a hard symmetry breaking, 
and furthermore no spontaneous breaking has been observed until now.

Hence, we consider a model with $SU(2)_L\otimes SU(2)_R\otimes U(1)_{B-L}$ gauge symmetry 
without introducing the parity or charge conjugation symmetry from the beginning. The parity is broken explicitly and the left-handed weak interactions are not related with the right-handed ones. Of course, the $SU(2)_R$ symmetry involves heavy vector bosons in order to be compatible with low energy phenomenology while at high energies there is no parity restoration at all.

Independently of the way in which parity is broken, right-handed neutrinos are included since neutrinos must be massive particles. Moreover, neutrinos may be Dirac or Majorana particles depending on the scalar representation of the model to be considered.  For instance, if active neutrinos were Majorana fermions and right-handed neutrinos were heavy, complex scalar triplets would have to be added and type-I seesaw  mechanism would be implemented~\cite{Mohapatra:1979ia}. On the contrary, if neutrinos were Dirac fermions~\cite{Senjanovic:1978ev} scalar bi-doublets and a doublet of $SU(2)_R$ would have to be considered and there would be no heavy right-handed neutrinos since they are related with the active ones in the Dirac field $\nu=\nu_L+\nu_R$ with the mass term $m_D\bar{\nu}\nu$. In this case the smallness of the neutrino masses need a fine tuning unless three bi-doublets are introduced~\cite{Chavez:2019yal}.  

As we assume explicit parity violation from the very begining: $g_L\not=g_R$, the left- and right- mixing matrices are different 
from each other, and also the scalar particle content is not constrained, since each right-handed scalar multiplet does not necessarily have left-handed partner. Only in the fermionic sector we introduce both left- and right-handed doublets, but 
this is not imposed in the scalar sector: In other words, we introduce only scalar doublets or triplets under $SU(2)_R$ without their respective counterparts under $SU(2)_L$. Then, from the point of view of the parity violation, this model is not better than the SM;
it just places the left- and right-handed components of the known fermions on the same footing, and also predicts new weak interactions. 

The structure of this paper is as follows. In Sec.~\ref{sec:model}, we describe the particle spectrum of the model, while in Sec.~\ref{sec:gauge},
we analyse the gauge boson sector by considering one scalar bi-doublet, $\Phi\sim(\mathbf{2}_L,\mathbf{2}_R,0)$, and one  doublet of $SU(2)_R$, $\chi_R\sim(\textbf{1}_L,\textbf{2}_R,+1)$. The fermion-vector boson interactions are considered 
in Sec.~\ref{sec:leptonsw} and, in Sec.~\ref{sec:yukawa} we deal with the lepton mass generation and Yukawa interactions with the scalar multiplets considered in Sec.~\ref{sec:gauge}; in this case, neutrinos are Dirac particles. The introduction of a scalar triplet like $\Delta_R\sim (\textbf{1}_L,\textbf{3}_R,2)$, could have been
considered, but it is not relevant for the main purpose of this paper. Phenomenological consequences and conclusions are discussed 
in Secs.~\ref{sec:feno} and \ref{sec:con}, respectively.  

\section{The model}
\label{sec:model}

The electroweak gauge symmetries of the model are $SU(2)_L\otimes SU(2)_R\otimes U(1)_{B-L}$ and $g_L\not=g_R$ at any energy scale. We omit the $SU(3)_C$ factor, because is similar to the SM. 
Hence, this model has three gauge couplings $g_L,g_R$ and $g_{BL}$ for each group. The electric charge operator is 
defined \cite{Marshak:1979fm} as usual $Q/e= T_{3L}+T_{3R}+(B-L)/2$.

The fermion sector consits of: leptons  which transform as $L^\prime_a =(\nu^\prime_a\;
l^\prime_a )^T_L\sim (\textbf{2}_L,\textbf{1}_R,-1)$ and  $R^\prime_a =(\nu^\prime_a\;
l^\prime_a )^T_R\sim (\textbf{1}_L,\textbf{2}_R,-1)$,
with $l=e,\mu,\tau$. Similarly, for quarks $Q^\prime_{iL}$ and $Q^\prime_{iR}$ with $(B-L)=1/3$. The primed states denote symmetry eigenstates.

The scalar sector consists in one or more bi-doublets transforming as $(\textbf{2}_L,\textbf{2}^*_R,0)$: 
\begin{equation}
\Phi = \left(
\begin{array}{cc}
\phi^{0} & \eta^{+}\\
\phi^{-} & \eta^{0}
\end{array}\right)
\label{higgs1}
\end{equation}
for generating the fermion masses, one doublet $\chi_R\sim(\textbf{1}_L,\textbf{2}_R,+1)$ if we want heavy right-handed neutrinos, 
and one triplet $\Delta_R\sim(\textbf{1}_L,\textbf{3}_R,+2)$ to implement the type I seesaw mechanism. Here, for simplicity, we consider,
 the case of just one bi-doublet $\Phi$, and one right-handed doublet $\chi_R=(\chi^+_R\;\chi^0_R)^T$
to break the gauge symmetry into the SM group, since these are enough to demostrate the main features of this type of models. 
A doublet $\chi_L\sim(\textbf{2}_L,\textbf{1}_R,+1)$ or a triplet $\Delta_L\sim(\textbf{3}_L,\textbf{1}_R,+2)$ could also be added as
inert scalars.

We write the vacuum expectation values (VEVs) as $\langle\Phi\rangle=\textrm{Diag}(ke^{i\delta}\;\;k^\prime)/\sqrt{2}$ and 
$\langle \chi_R\rangle =(0\;v_R)^T/\sqrt{2}$ since one of the phases in $\Phi$ and the phase in $\chi_R$ can be always eliminated by 
unitary unimodular transformations. Complex VEVs implies 
spontaneous~$C\!P$~violation~\cite{Senjanovic:1978ev,Basecq:1985sx}.

\section{Gauge boson mass eigenstates}
\label{sec:gauge}

The covariant derivative for the bi-doublet $\Phi$ and the doublet $\chi_R$ are given by: 
\begin{eqnarray}
\mathcal{D}_{\mu }\Phi&=&\partial _{\mu }\Phi+i\left[g_L\frac{\vec{\tau}}{2}\cdot\vec{W}_{L\mu}\,\Phi-g_R\Phi \,\frac{\vec{\tau}}{2}\cdot\vec{W}_{R\mu}\right],\quad (a)\nonumber \\
\mathcal{D}_{\mu }\chi_R&=&\left(\partial _{\mu }+ig_R \frac{\vec{\tau}}{2}\cdot \vec{W}_{ R\mu}-ig_{BL} B_{\mu }\right)\chi_R,\quad \quad \quad \;\,(b)
\label{dc1}
\end{eqnarray}
where $g_L\not=g_R$ and the Lagrangian is $\mathcal{L}_{XW}=(\mathcal{D}^{\mu }X)^\dagger (\mathcal{D}_{\mu }X) $ with $X=\Phi,\chi_R$.
Using (\ref{dc1}) in $\mathcal{L}_{XW}$, the charged-boson mass matrix in the basis $(W^+_L\,W^+_R)^T$ has the form:
\begin{eqnarray}
\mathcal{M}^2_{CB}=\frac{g^2_Lv^2_R}{4}\left(
\begin{array}{cc}
x & -2\epsilon ze^{i\alpha}\\
-2\epsilon ze^{-i\alpha} &1+x 
\end{array}
\right),
\label{mcb1}
\end{eqnarray}
where $x=K^2/v^2_R$, $z=\bar{K}^2/v ^2_R$ and
$K^{2}=k^{2}+k^{\prime 2}$, $\bar{K}^2=\vert k\vert \vert k^{\prime}\vert$.
Hereafter we will consider $\alpha=0$ i.e., the VEVs are real numbers. The masses of the charged gauge bosons are given by:
\begin{eqnarray}
M^2_{W_1}=\frac{g^2_Lv^2_R}{8}\left((1+\epsilon^2) x +\epsilon^2\,- \sqrt{\Delta}\right),\quad
M^2_{W_2}=\frac{g^2_Lv^2_R}{8}\left( (1+\epsilon^2)x +\epsilon^2+ \sqrt{\Delta}\right),  
\label{mcb2}
\end{eqnarray}
where
\begin{equation} 
\Delta= 4\,\epsilon^2 x^2 +\left[\epsilon^2+(\epsilon^2-1 )x \right]^2,\quad\epsilon=g_R/g_L,\quad x=K^2/v^2_R.
\label{delta}
\end{equation}
	
The mass eigenstates in terms of the symmetry eigenstates are as follows:
\begin{equation}
W^{+}_{L\mu}=c_{\xi}\,W^{+}_{1\mu}-s_{\xi}\,W^{+}_{2\mu},\quad
W^{+}_{R\mu}=s_{\xi}\,W^{+}_{1\mu}+c_{\xi}\,W^{+}_{2\mu},
\label{mcb3}
\end{equation}
where
\begin{equation}
s_{\xi}=\frac{4\,g_Lg_R\,\bar{K}^2 }{\sqrt{16g^2_Lg^2_R\bar{K}^4+Y^2} },\quad
c_{\xi}=\frac{Y}{\sqrt{16g^2_Lg^2_R\bar{K}^4+Y^2} },
\label{bcm4}
\end{equation}
and $$Y=K^2(g^2_R-g^2_L)+g^2_R v^2_R+\sqrt{\Delta^{\prime}},\,\,\mbox{with} \,\,
\Delta^{\prime}=16g^2_Lg^2_R\bar{K}^4+[g^2_R v^2_R+K^2(g^2_R-g^2_L)]^2,$$
Assuming $v^2_R \gg \vert k\vert^2,\vert k^\prime\vert^2$:
\begin{equation}
M^2_{W_1}\approx \frac{g^2_L}{4}K^2 ,\quad
M^2_{W_2}\approx\frac{g^2_R}{4} v^2_R,\quad \xi\approx \pm \frac{1}{\epsilon}\frac{\bar{K}^2}{v^2_R}.
\label{mcb5}
\end{equation}
 
We note that $M^2_{W_2}\gg M^2_{W_1}$, and in this limit, we obtain $W_L\approx W_1$ and $W_R\approx W_2$.

Also, from (\ref{dc1}) and $\mathcal{L}_{XW}$, the neutral vector boson mass matrix in the basis $(W_{3L},W_{3R},B)^T$ is:
\begin{eqnarray}
\mathcal{M}^2_{NB}=\frac{g^2_Lv^2_R}{4}\left(
\begin{array}{ccc}
x &-\epsilon x&0\\
-\epsilon x &\epsilon^2(1+x)& -\epsilon\,\delta_L  \\
0 &-\epsilon\,\delta_L  & \delta^2_L
\end{array}
\right),
\label{mnb1}
\end{eqnarray}
where $\epsilon$ and $x$ are defined in Eq.~(\ref{delta}). The matrix above has the following eigenvalues:
\begin{eqnarray}
M^2_A&=&0,\nonumber \\
M^2_{Z_1}&=&\frac{g^2_Lv^2_R}{8}\left[(1+\epsilon^2)x+\delta^2_L +\epsilon^2-\sqrt{\Delta_N}\right],\nonumber \\
M^2_{Z_2}&=&\frac{g^2_Lv^2_R}{8}\left[(1+\epsilon^2 )x+\delta^2_L +\epsilon^2 +\sqrt{\Delta_N}\right],
\label{mnb2}
\end{eqnarray}
Where: 
\begin{eqnarray}
&&\delta_L=\frac{g_{BL}}{g_L},\quad \delta_R=\frac{g_{BL}}{g_R}\nonumber \\&&
\Delta_N=\left(1+\epsilon^2\right)^2\,x^2  +2  [\epsilon^2(\epsilon^2-1) -\delta^2_L(\epsilon^2+1) ]x + (\delta^2_L +\epsilon^2)^2,
\label{deltapp}
\end{eqnarray}
with $\epsilon$ defined in Eq.~(\ref{delta}).

Again, making $v^2_R \gg \vert k\vert^2,\vert k^\prime\vert^2$:
\begin{equation} 
M^2_{Z_1}\approx\frac{g^2_LK^2}{4}
\left( \frac{\epsilon^2+(1+\epsilon^2)\delta^2_L}{\epsilon^2+\delta^2_L}\right),\quad
M^2_{Z_2}\approx\frac{g^2_Lv^2_R}{4}(\epsilon^2+\delta^2_L ).
\label{mnb4}
\end{equation}
Notice that $M^2_{Z_2}\gg M^2_{Z_1}$ and $M_{Z_2}>M_{W_2}$. 
	
After diagonalizing the neutral mass matrix by an orthogonal matrix, $O^T\mathcal{M}^2_{NB}O=\hat{M}=(0,M^2_{Z_1},M^2_{Z_2})$, 
we obtain the following symmetry eigenstates as a function of mass eigenstates:
\begin{eqnarray}
\left(\begin{array}{c}
W^{L}_{3\mu}\\
W^{R}_{3\mu}\\
B_{\mu}
\end{array}\right)=\left(\begin{array}{ccc}
\frac{1}{N_1} & \frac{1}{N_2}& \frac{1}{N_3}\\
\frac{1}{N_1\epsilon} & -\frac{a_1}{N_2}&-\frac{a_2}{N_3}\\
\frac{1}{N_1\delta_L}&-\frac{b_1}{N_2}& -\frac{b_2}{N_3}
\end{array}\right)\left(\begin{array}{c}
A^{\mu}\\
Z^{\mu}_1\\
Z^{\mu}_2
\end{array}\right),
\label{eigen}
\end{eqnarray}
where:                                                                                                                                             
\begin{eqnarray}
2a_{1}&=& \epsilon-\frac{1}{\epsilon}-(\delta_L\delta_R+\epsilon)\frac{v^2_R}{K^2}-\sqrt{\frac{\Delta_N}{ K^4}},\hspace{0.5cm}
\;\; b_{1}=\delta_L\left[1+\frac{1}{\epsilon^2}-(\delta^2_R-1)\frac{v^2_R}{K^2}+\sqrt{\frac{\Delta_N}{K^4}}\right],
\nonumber \\
2a_{2}&=&\epsilon-\frac{1}{\epsilon}+(\delta_L\delta_R +\epsilon)\frac{v^2_R}{K^2}+\sqrt{\frac{\Delta_N}{K^4}} ,\hspace{0.9cm}
b_{2}=\delta_L\left[1+\frac{1}{\epsilon^2}-(\delta^2_R-1)\frac{v^2_R}{K^2}-\sqrt{\frac{\Delta_N}{K^4}}\right],\nonumber \\&&
N_1=\sqrt{1+\epsilon^{-2}+\delta_L^{-2}} ,\hspace{0.9cm}
N_2=\sqrt{1+a^2_1+b^2_1},\hspace{0.9cm}
N_3=\sqrt{1+a^2_2+b^2_2},
\label{mnb5}
\end{eqnarray}
$\delta_{L(R)}$ and $\Delta_N$ are given in Eq.~(\ref{deltapp}). The eigenstates are normalized and we stress that these are exact results. 

\section{Lepton-vector boson interactions}
\label{sec:leptonsw}

When $g_L\not= g_R$ the covariant derivatives in the lepton sector are
\begin{eqnarray}
&&\mathcal{D}^L_\mu = \partial_\mu+i\frac{g_L}{2}\,\vec{\tau}\cdot\vec{W}_{L\mu}+iY_L\frac{g_{BL}}{2}B_\mu,\;
\mathcal{D}^R_\mu =\partial_\mu+i\frac{g_R}{2}\,\vec{\tau}\cdot\vec{W}_{R\mu}+iY_R\frac{g_{BL}}{2}B_\mu, 
\label{dc2}
\end{eqnarray} 
where $\mathcal{D}^L_\mu$ and $\mathcal{D}^R_\mu$ acts on left- and right-handed lepton doublets, similarly for quarks and $Y_{L,R}=-1(1/3)$ for leptons(quarks).

\subsection{Charged interactions}
\label{subsec:ci} 

When derivatives of Eq.~(\ref{dc2}) are applied in the quark sector, we have,
in the mass eigenstate basis,
\begin{eqnarray}
\mathcal{L}^q_W&=&-\frac{1}{2}\left[e^{i\phi_q}g_L J^{q\mu}_L  W^+_{L\mu}+g_R J^{q\mu}_R   W^+_{R\mu} \right]+H.c.\nonumber \\&=&
-\frac{1}{2}\left[(e^{i\phi_q}c_\xi g_L J^{q\mu}_L +s_\xi g_RJ^{q\mu}_R ) W^+_{1\mu}-
(e^{i\phi_q}s_\xi g_LJ^{q\mu}_L -c_\xi g_RJ^{q\mu}_R ) W^+_{2\mu} \right]\!+\!H.c.,
\label{ci1}
\end{eqnarray}
where we have used Eq.~(\ref{mcb3}). Recall that 
$J^{q\mu}_L=\bar{u}_L\gamma^\mu V^L_{CKM} d_L$ and $J^{q\mu}_R=\bar{u}_R \gamma^\mu \tilde{V}^R_{CKM}d_R$ where 
$\tilde{V}^R_{CKM}=K^{u*}V^R_{CKM} K^d$. $V^L_{CKM}$ is the same as in the left-handed sector of the SM with three angles
and one physical phase (the $K^u$, $K^d$ diagonal matrices are shown below).  We obtain, as usual, $V^{L,R}_{CKM}=V^{u\dagger }_{L,R}V^d_{L,R}$. 
Notice that, unlike the manifest parity-invariant case (in which the fermion mass matrices are Hermitian, implying that $V^d_L=V^ d_R\equiv V^d$, 
$V^ u_L=V^ u_R\equiv V^u$ and $V^ L_{CKM}=V^ R_{CKM}=V^ {u\dagger}V^ d=V_{CKM}$), in this work, we have $V^L_{CKM}\not=V^R_{CKM}$.
This has important consequences for the number of  $CP$ violating physical phases. 
In fact, the phases $\phi_q$ in Eq.~(\ref{ci1}) cannot be absorbed in the fields and are le\-gi\-ti\-ma\-te physical phases. 

On the other hand, it is known that there are $n(n+1)/2$ phases in $n\times n$ unitary matrices. In such a matrix between two spinors $\bar{u}_L\gamma^\mu V d_L$, one phase is global and $2n-1$ phases can be absorbed
in $2n-1$ mass eigenstates left-handed fields, say $u_L\to K^u u_L$ and $d_L \to K^d d_L$, in the quark 
sector, where $K^u=\textrm{Diag}(e^{ia_{u_1}},e^{ia_{u_2}}, \cdots,e^{ia_{u_n}}  )$ and 
$K^d=\textrm{Diag}(1,e^{ia_{d_2}}, \cdots,e^{ia_{d_n}})$. Hence, the number of physical phases is given by $(n-2)(n-1)/2$.
This re-phasing is done in the mass eigenstate basis and they are absorbed in the mass term and neutral currents (in the SM neutral 
currents, they are coupled with $Z$ and the Higgs $h$ are diagonal) by the mass eigenstates of the right-handed fields to keep the masses real, i.e., 
$u_R\to K^u u_R$, $d_R\to K^d d_R$. Then $V\to K^{u*}V K^d=e^{i(a_{d_j}-a_{u_k})}V_{k j}\equiv V^L_{CKM}$.
 However, this rephasing does not work if there are flavour changing neutral 
currents (FCNC) and/or, as in the present case, there are more charged currents. In fact, in the present case, there are Flavor Chanching Neutral Currents (FCNC) in the Higgs 
sector (see Sec.~\ref{sec:yukawa}) and it is not longer possible to absorbe the phases of the left-handed fields into the right-handed ones. 
Furthermore, if we redefine the phases of the left-handed current, in order to have these interactions (since they are present in the SM), the phases will 
appear in the right-handed currents. Hence, the $V^R_{CKM}$ mixing matrix has $n(n+1)/2$ physical phases, and there is also a phase difference between the left- and right-handed currents, $\phi_q$. 

Similarly, from Eq.~(\ref{dc2}), in the leptonic sector, and in the mass eigenstate basis, with phase matrices defined $K^\nu$ and $K^l$ defined as in the quark sector, we have
\begin{eqnarray}
\mathcal{L}^l_W=
-\frac{1}{2}\left[e^{i\phi_l}g_L\left(c_\xi  W^+_{1\mu} -s_\xi W^+_{2\mu} \right) J^{l\mu}_L +g_R\left(s_\xi  W^+_{1\mu} +c_\xi  W^+_{2\mu}\right)J^{l\mu}_R \right]+H.c.
\label{ci2}
\end{eqnarray}
where $J^{l\mu}_L=\bar{l}_L\gamma^\mu V^L_{PMNS} \nu_L$ and $J^{l\mu}_R=\bar{l}_R \gamma^\mu \tilde{V}^R_{PMNS}\nu_R$, with $\tilde{V}^ R_{PMNS}=K^ {l*}V^ R_{PMNS}K^\nu$.
The physical phases in the lepton sector are, for the same reason as before, $\phi_l$, $K^\nu$ and $K^l$. 
The matrix $V^L_{PMNS}$ has, as usual, one phase in the Dirac case, or three phases in the Majorana case, but $V^R_{PMNS}$ has $n(n+1)/2$ physical phases.
We note from Eqs.~(\ref{ci1}) and (\ref{ci2}) that the right-handed current 
that is coupled with $W^\pm_1\approx W^\pm$ is suppressed by both $s_\xi$ and $g_R$ and those currents
coupled with $W^\pm_2$ are suppressed by $g_R$ and the mass of the $W^\pm_2\approx W^\pm_R$. 

\subsection{Neutral interactions}
\label{subsec:ni}

The neutral interactions of a fermion $i$ with the $Z_{1\mu}$ and $Z_{2\mu}$ neutral bosons are as follows:
\begin{equation}
\mathcal{L}_{NC}=-\frac{g_L}{2\cos\theta}\sum_i\bar{\psi}_i\gamma^\mu[(g^i_V-g^i_A\gamma^5)Z_{1\mu}+(f^i_V-f^i_A\gamma^5)Z_{2\mu}]\psi_i
\label{nc1}
\end{equation} 
We define 
\begin{eqnarray}
g^i_V=\frac{1}{2}(a^i_L+a^i_R),\;\; g^i_A=\frac{1}{2}(a^i_L-a^i_R),\quad
f^i_V=\frac{1}{2}(\mathcal{A}^i_L+\mathcal{A}^i_R),\;\; f^i_A=\frac{1}{2}(\mathcal{A}^i_L-\mathcal{A}^i_R),
\label{defas}
\end{eqnarray}
where $a^i_L$ and $a^i_R$ ($i$: $1$, $2$ and $3$) are couplings of the left- and right-handed components of a fermion $\psi_i$. Similarly, we define the couplings $\mathcal{A}^i_{L,R}$.

Using Eq.~(\ref{eigen}) we obtain for leptons:
\begin{eqnarray} 
&& a^\nu_L=\frac{1}{N_2}(1-\delta_L b_1),\hspace{1.2cm}
a^\nu_R=\frac{\epsilon}{N_2}(a_1-\delta_R  b_1) ,\nonumber \\ &&
a^l_L=-\frac{1}{N_2}(1+\delta_L  b_1),\hspace{0.9cm}
a^l_R=-\frac{\epsilon}{N_2} (a_1+\delta_R b_1).
\label{alarl}
\end{eqnarray}
or
\begin{eqnarray} 
&& g^\nu_V=\frac{1}{2N_2}(1+\epsilon\,a_1-2 b_1),\hspace{0.7cm}
g^\nu_A=\frac{1}{2N_2}(1-\epsilon \,a_1) ,\nonumber \\ &&
g^l_V=-\frac{1}{2N_2}(1+\epsilon\,a_1+2 b_1),\hspace{0.4cm}
g^l_A=-\frac{1}{2N_2}(1-\epsilon\,a_1),
\label{gvgal}
\end{eqnarray}
and, for quarks
\begin{eqnarray} 
&& a^u_L=\frac{1}{N_2}\left(1+\frac{1}{6}\delta_L b_1\right),\hspace{0.9cm}
a^u_R= \frac{\epsilon}{N_2}\left(a_1+\frac{1}{6}\delta_R b_1\right),\nonumber \\ &&
a^d_L= \frac{1}{N_2}\left(-1+\frac{1}{6}\delta_L b_1\right),\hspace{0.6cm}
a^d_R=\frac{\epsilon}{N_2}\left(-a_1+\frac{1}{6}\delta_R b_1\right).
\label{alarq}
\end{eqnarray}
or
\begin{eqnarray} 
&& g^u_V=\frac{1}{2N_2}\left(1+\epsilon\,a_1+\frac{1}{3}b_1\right),\quad
g^u_A=\frac{1}{2N_2}\left(1-\epsilon\,a_1 \right), \nonumber \\ &&
g^d_V= -\frac{1}{2N_2}\left(1+\epsilon\,a_1 -\frac{1}{3}b_1  \right),\;
g^d_A=\frac{1}{2N_2}\left(-1+\epsilon\,a_1 \right).
\label{gvgaq}
\end{eqnarray}

All these results are exact. If we expand up to $\mathcal{O}(1/v^2_R)$ we have for neutrinos
\begin{eqnarray}
a^{\nu}_L&\approx&1+\frac{\delta^2_R}{(1+\delta^2_R)^2}\,x,\hspace{1.9cm}
a^{\nu}_R\,\approx\,\frac{1}{1+\delta^2_R}\,x,\nonumber\\
g^{\nu}_V&\approx&\frac{1}{2}\left(1+ \frac{1+2\delta^2_R}{(1+\delta^2_R)^2}\,x\right),\quad
g^{\nu}_A\approx\,\frac{1}{2}\left(1-\frac{1}{(1+\delta^2_R)^2}\,x\right),
\label{nusapp}
\end{eqnarray}
and
 \begin{eqnarray}
a^{\ell}_L&\approx&\frac{\delta^2_L-\delta^2_R-1}{\delta^2_L+\delta^2_R+1}+\frac{\delta^2_R}{(1+\delta^2_R)^2}\,x,\hspace{1.4cm}
a^{\ell}_R\approx\frac{2\,\delta^2_L}{\delta^2_L+\delta^2_R+1}+\frac{\delta^2_R-1}{(1+\delta^2_R)^2}\,x,\nonumber\\
g^{\ell}_V&\approx&\frac{1}{2}\left(\frac{3\delta^2_L-\delta^2_R}{\delta^2_L+\delta^2_R+1}+\frac{2\delta^2_R-1}{(1+\delta^2_R)^2}
\,x\right),\hspace{0.6cm}
g^{\ell}_A\approx \frac{1}{2}\left(-1+\frac{1}{(1+\delta^2_R)^2}\,x\right),
\label{clapp}
\end{eqnarray}
for charged leptons. 

Notice that when $v_R\to\infty(x\to 0)$ and $\delta_L=\delta_R$ we obtain,
\begin{eqnarray}
g^\nu_V,g^\nu_A\to \frac{1}{2},\hspace{1cm} g^l_V\to -\frac{1}{2}+2s^2_{\theta},\,\,\,g^l_A\,\,\to\,\, -1/2.
\label{gnun}
\end{eqnarray}
The same happens with the coefficients of the quarks in that limit. 

The $f$'s in Eq.~(\ref{nc1}) are obtained by making $N_2\to N_3$, $\,\,a_1\to a_2\,\,$ and $\,\,b_1\to b_2$ in Eqs.~(\ref{alarl})--(\ref{gvgaq}). We
do not write the $g$'s and $f$'s of the quark sector.

\subsection{Electromagnetic interactions}
\label{sec:emi}

From the projection on the photon field in Eq.~(\ref{eigen}), using the covariant derivative in Eq.~(\ref{dc2}), we obtain the electric charge of a charged fermions, $f$:
\begin{equation}
eQ_{f_L}=eQ_{f_R}\equiv eQ_f=\frac{g_L}{N_1},
\label{carga1}
\end{equation}
where $Q_l=1$ for positive leptons and $Q_u=2/3,Q_d=-1/3$ for quarks (in units of the positron charge), and $Q_\nu=0$ for neutrinos. In fact, using Eq.~(\ref{eigen}) in  the interaction terms with the photon, we obtain an exact zero for the neutrino's electric charge:
$Q_{\nu L(R)}=g_Lg_Rg_{BL}-g_{BL}g_Rg_L\equiv 0$ which also implies that $Q_l=1$. 

From Eq.~(\ref{carga1}) and $N_1$ given in Eq.~(\ref{mnb5}), we obtain
\begin{equation}
\frac{1}{e^2}=\frac{1}{g^2_L}+\frac{1}{g^2_R}+\frac{1}{g^2_{BL}}.
\label{relacao}
\end{equation}
Note that $g_L,g_{BL}$ may be related to SM $g,g_Y$ coupling constants only at a given energy, say $\mu_m$:
\begin{equation}
g(\mu_m)=g_L(\mu_m),\quad \mbox{and}\quad\quad\frac{1}{g^2_Y(\mu_m)}=\frac{1}{g^2_R(\mu_m)}+\frac{1}{g^2_{BL}(\mu_m)}.
\label{matching}
\end{equation}
For energies $\mu\not=\mu_m$ the couplings $g_L,g_R$ have different running and no longer equal to each other,
even in the left-right symmetric model~\cite{Chang:1983fu,Chang:1984uy}. 
From Eq.~(\ref{matching}) defining $\tan\phi=g_Y/g_R$, we obtain 
\begin{equation}
\tan^2\phi=\frac{\delta^2_R}{1+\delta^2_R},
\label{matching2}
\end{equation}
where $\delta_R$ was defined in Eq.~(\ref{deltapp}), or by rewriting the last equation, we obtain
\begin{equation}
\delta^2_R=\frac{s^2_\phi}{1-2s^2_\phi}.
\label{polo1}
\end{equation}	
Note that when $s^2_\phi=1/2$ there is a Landau-like pole in $g_{BL}$. Notwithstanding, Eq.~(\ref{polo1}) is valid at any energy scale, since it does not involve an SM parameter, i.e., it is not a matching condition, but just a definition. 

This is consistent with the two independent parameters called $\varepsilon$ R models (in our notation $\varepsilon=\epsilon^{-1}$) and $\theta^\prime$ in Ref.~\cite{Kokado:2015iya}. The difference between our calculations and those in Ref.~\cite{Kokado:2015iya} is that we explicitly diagonize the mass square matrix of the neutral vector boson given in Eq.~(\ref{eigen}) and those authors of \cite{Kokado:2015iya} assume at the very start that the relation among $(W_{3L}\,,W_{3R},\,B)$ and $(A,Z,Z^\prime)$ is a $3\times3$ arbitrary orthogonal matrix with three angles, they then impose constraints in order to obtain the correct electric charge. Moreover, in \cite{Kokado:2015iya} the condition $g_L\not=g_R$ arises because of the different running of these couplings' constants, since $g_L=g_R$ is valid only at a given energy. 


The fact that the fermion electric charge imposes constraints on an arbitrary orthogonal matrix relating the symmetry
eigenstates $W_3,W_R,B$ with mass eigenstates $A,Z_1,Z_2$ was noted in the context of left-right symmetric models in 
Ref.~\cite{Kokado:2015iya} and in secluded $U(1)$ models in Ref.~\cite{Fortes:2017kca}. In fact, from Eq.~(\ref{carga1}) we can define two angles $e=g\sin\kappa$ and $e=g_{BL}\cos\zeta$ where
\begin{equation}
\sin\kappa=\frac{g_Rg_{BL}}{\sqrt{(g^2_R+g^2_L)g^2_{BL}+g^2_Lg^2_R}},\quad
\cos\zeta=\frac{g_Rg_L}{\sqrt{(g^2_R+g^2_L)g^2_{BL}+g^2_Lg^2_R}},
\label{2angles}
\end{equation} 
and only when $g_L=g_R\equiv g$ can the matrix in Eq.~(\ref{eigen}) be parametrized by only one angle, say $\theta$, and
\begin{equation}
e=g\sin\theta, \quad e=g_{BL}\sqrt{\cos2\theta}
\label{charge} 
\end{equation}

In fact, only when $g_L=g_R$ is valid at any energy (as is usual assumed), there is only an angle, say $\theta$, $\theta\to \theta_W$ and the pole in Eq.~(\ref{polo1}) arises if $s^2_W=1/2$. This is, in fact, the usual assumption in left-right symmetrics models; the invariance under a generalized parity symmetry implies
$g_L=g_R\equiv g$, from which, using Eq.~(\ref{relacao}) we can obtain
\begin{equation}
\frac{1}{e^2}=\frac{2}{g^2}+\frac{1}{g^2_{BL}},
\label{carga5}
\end{equation}

However, we stress that in the present model just the two angles defined in Eq.~(\ref{2angles}) are the only ones.

\section{Yukawa interactions}
\label{sec:yukawa}

The Yukawa interactions in the quark and lepton sector are given by
\begin{eqnarray}
-\mathcal{L}^Y&=&\overline{L^\prime_a}\,(G_{lab}\Phi+ F_{lab}\tilde{\Phi})\, R^\prime_b+\overline{R^\prime_a}\,(G^\dagger_{lab}\Phi^\dagger+ F^\dagger_{lab}\tilde{\Phi}^\dagger)\,L^\prime_b\nonumber \\&+&\overline{Q^\prime_{iL}}\,(G_{qij}\Phi+ F_{qij}\tilde{\Phi})\, Q^\prime_{jR}+\overline{Q^\prime_{iR}}\,(G^\dagger_{qij}\Phi^\dagger+ F^\dagger_{qij}\tilde{\Phi}^\dagger)\,Q^\prime_{jL},
\label{yukawa1}
\end{eqnarray}
where $L^\prime$ and $R^\prime$ are defined in Sec.~\ref{sec:model}, $G_{l,q}$ and $F_{l,q}$ are arbitrary $3\times3$ matrices in the flavor space.
Notice that since we are not imposing a generalized parity, these matrices are not Hermitian as in the left-right symmetric case. Hence they are diagonalized by bi-unitary transformations $V^{\nu,l}_{L,R}$ and $V^{u,d}_{L,R}$. In addition, the VEVs may be complex numbers.

\begin{eqnarray}
-\mathcal{L}^Y_l&=&\bar{\nu}_LV^{\nu\dagger}_L[ (G_l\phi^0+F_l\eta^{0*})V^\nu_R\,\nu_R+(G_l\eta^+ -F_l\phi^+)V^l_Rl_R]\nonumber \\&+&
\bar{l}_LV^{l\dagger}_L[(G_l\phi^- -F_l\eta^-)V^\nu_R\, \nu_R+(G_l\eta^0+F_l\phi^{0*})V^l_Rl_R]+H.c.
\label{yukawa2}
\end{eqnarray} 

In the quark sector we have 
\begin{eqnarray}
-\mathcal{L}^Y_q&=&\bar{u}_LV^{u\dagger}_L[ (G_q\phi^0+F_q\eta^{0*})V^u_R\,u_R+(G_q\eta^+ -F_q\phi^+)V^d_Rd_R]\nonumber \\&+&
\bar{d}_LV^{d\dagger}_L[(G_q\phi^- -F_q\eta^-)V^u_R\, u_R+(G_q\eta^0+F_q\phi^{0*})V^d_Rd_R]+H.c.
\label{yukawa3}
\end{eqnarray} 
where the scalar are still symmetry eigenstates. We have omitted flavour indices. 
Notice that there are flavour changing neutral interactions with the neutral scalars 
in both the lepton and quark sectors. Of course, more scalar bi-doublets may be introduced as in Ref.~\cite{Chavez:2019yal}.

\section{Phenomenological consequences}
\label{sec:feno}

Here, we have presented a model with left and right gauge groups in which parity is broken explicitly. 
It is not the objective of the this work to make a detailed analysis of the phenomenology of this model nor to compare it 
with other models with manifest, or pseudo-manifest, parity invariance. However, we will briefly discuss some processes that, although
they may occur in other models with left and right gauge symmetries, have different numerical results, since the parameters 
related to the right-handed fermionic sector are completely arbitraries and also the scalar sector in the present model is different from that
in other models: we do not require that the scalar sector must be symmetric under a generalized parity, charge conjugation, or any other discrete symmetry. 

For instance, there is no left-handed doublets or triplets, nor a scalar singlet required to break parity first, other than the gauge symmetries~\cite{Chang:1983fu,Chang:1984uy}.
In the particular model considered, only a doublet was introduced under $SU(2)_R$, besides at least one bi-doublet. In this case, if a $SU(2)_L$ doublet were introduced, 
it could be an inert scalar since it does not couple with any fermion at all; and if $\langle \chi_L\rangle=0$, its neutral component
does not contribute to the spontaneous breakdown of the $SU(2)_L$ symmetry. On the other hand, if a left-handed triplet $\Delta_L$ were to be introduced then an 
appropriate discrete symmetry could be added in order to avoid couplings such as $\overline{L^c_l}\Delta_LL_l$. If this were the case, the triplet might be inert with $\langle \Delta^0_L\rangle=0$.

In this context, the fact that weak interactions are parity asymmetric at low energies is because right-handed currents are suppressed by the great mass of the respective vector bosons and/or by the small size of the gauge coupling $g_R$, i.e., $g_R\ll g_L$. As usual, the heaviness of the vector bosons related to the right-handed symmetry is due to a VEV larger than the electroweak scale.  

In this model there is universality in quark and lepton sectors, but the universality in the left-handed interactions is different from that  
 of the right-handed ones because $g_R \not=g_L$, $V^L_{CKM}\not= V^R_{CKM}$, $V^L_{PMNS}\not= V^R_{PMNS}$, hence $W^-_R\to e^-_R(\nu_R)^c$ has no relation to $W^-_L\to e^-_L(\nu_L)^c$. Moreover, there are no
flavour changing neutral currents in the vector interactions; however, for a fermion $i$, the couplings with the $Z_1\approx Z$ in the limit $v_R\to \infty$ are equal to the tree level values, as in the SM: $g^i_{V,A}= g^i_{(V,A)SM}$, and this has been noted in Eq.~(\ref{gnun}). 
If neutrinos are Dirac particles, right-handed neutrinos are just components of a Dirac spinor. In this case, there are no heavy right-handed 
neutrinos and most phenomenological studies do not apply, at least, not in a straightforward way. In fact, in the model shown above, only Dirac neutrinos have been considered, since no scalar triplets are introduced. However, we can compare some processes which are important when neutrinos are Dirac or Majorana particles. For instance,
\begin{eqnarray}
&& qq^\prime\to W^+_R\to \nu_Rl^+_{1L}\to W^-_R l^+_{1L} l^+_{2L} \to l^+_{1L}l^+_{2L}+jets(leptons), \;\;(a)\;\;\; \textrm{M}
\nonumber \\&&
qq^\prime\to W^+_L\to \nu_Ll^+_{1R}\to W^-_L l^+_{1R}l^+_{2R}\to l^+_{1R}l^+_{2R}+jets(leptons),\;\;(b)\;\;\; \textrm{M}\nonumber \\ &&
qq^\prime\to W^+_R\to \nu_R l^+_{1L}\to W^+_R l^+_{1L}l^-_{2R}\to 
l^+_{1L}l^-_{2R}+jets(leptons),\;\,\,(c)\;\;\; \textrm{D,M}\nonumber \\ &&
qq^\prime\to W^+_L\to \nu_Ll^+_{1R}\to W^+_L l^+_{1R} l^-_{2L} \to l^+_{1R}l^-_{2L}+jets(leptons),\;\;(d)\;\;\; \textrm{D,M}
\nonumber \\ &&
qq^\prime\to W^+_R\to \nu_R l^+_{1L}\to W^+_L l^+_{1L} l^-_{2L}\to l^+_{1L}l^-_{2L}+jets(leptons),\;\;(e)\;\;\; \textrm{D}\nonumber \\ &&
qq^\prime\to W^+_L\to \nu_L l^+_{1R}\to W^+_R l^+_{1R} l^-_{2R}\to l^+_{1R}l^-_{2R}+jets(leptons),\;\;(f)\;\;\; \textrm{D}
\label{ksp}
\end{eqnarray}  

\begin{center}
\begin{figure}[!h]
\includegraphics[width=10cm]{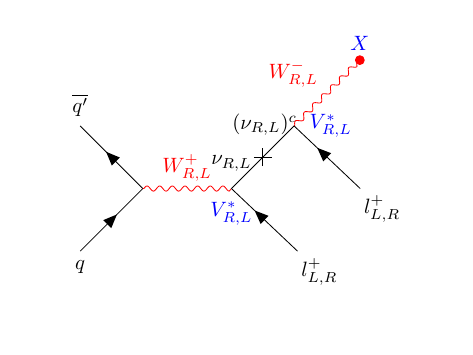}
\caption{Processes in Eqs.~(\ref{ksp}(a)) and (\ref{ksp}(b)) that are induced by pure Majorana neutrinos. When $X$ are jets, this is the Keung-Senjanovic (KS) process~\cite{Keung:1983uu}. However, they also may be leptons~\cite{Helo:2018rll}. We only explicitly show the mixing matrices in the lepton sector. In this figure, for the sake of simplicity, $V_{L,R}$ denotes the left- and right-handed mixing matrices, $V^L_{PMNS}$ and $V^R_{PMNS}$, respectively.}
\label{fig:fig1}
\end{figure}
\end{center}
\begin{center}
\begin{figure}[!h]
\includegraphics[width=10cm]{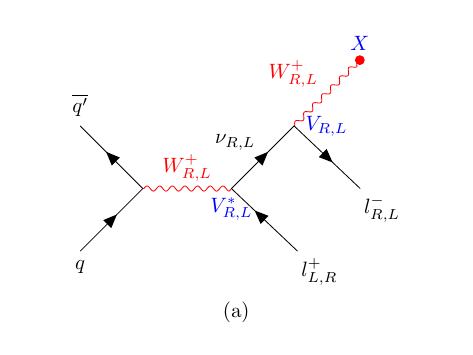}
\includegraphics[width=10cm]{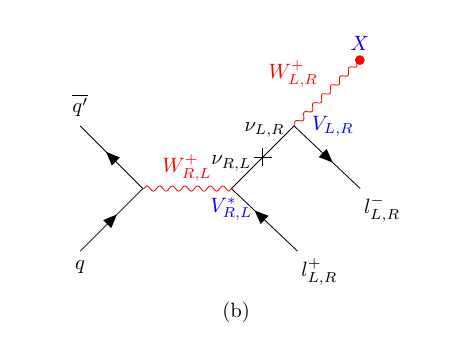}
\caption{In Fig.~2a we show a processes which occur with pure Dirac or purely Majorana neutrinos. See processes in (\ref{ksp}(c)) and (\ref{ksp}(d)). In Fig.~2b the process may only be induced by Dirac neutrinos. As in Fig.~\ref{fig:fig1} the red dot $X$ denotes jets, or leptons; and we only explicitly show the mixing matrices in the lepton sector. The notation for the mixing matrixes $V_L$ and $V_R$ is as shown in Fig.~\ref{fig:fig1}.}
\label{fig:fig2}
\end{figure}
\end{center}

If neutrinos are pure Majorana particles (with no right-handed neutrinos), processes $(a)$ and $(b)$ shown in Fig.~\ref{fig:fig1} occur. 
Processes such as those in $(c)$ and $(d)$ shown in Fig.~\ref{fig:fig2}(a) may occur when neutrinos are Dirac fermions, i.e. only scalar bi-doublets and one doublet $\chi_R$ are introduced; or if neutrinos are Majorana with with right-handed neutrinos, they are induced by a Dirac mass term i.e., when bi-doublet and a trilplet $\chi_R$ are introduced. Finally, processes such as $(e)$ and $(f)$, shown in Fig.~\ref{fig:fig2}(b) occur in both cases, i.e, pure Majorana or pure Dirac.

In the literature, only the 
case in Eq.~(\ref{ksp}a) has been considered, see, for instance~\cite{Ng:2015hba,Ruiz:2017nip}. Without going into the details of the processes given in Eq.~(\ref{ksp}), we note the following: In the Majorana case, Eq.~(\ref{ksp}-a) and (\ref{ksp}-b), the amplitude is proportional to $V^{R(L)*}_{PMNS}M^RV^{R(L)}_{PMNS}$, hence this case is favored by heavy neutrinos; while in $(c)$ and $(d)$ for light neutrinos, the amplitudes are  proportional to $\sum V^*_{ab} V_{bc}\sim\delta_{ac}$ ($V=V^R_{PMNS}$ or $V^L_{PMNS}$) and the charged lepton are the same in both vertices; or, if neutrinos are heavy, the amplitude is proportional to $V^\dagger M_R^{-1}V$ ($V=V^R_{PMNS}$ or $V^L_{PMNS}$) and are suppressed; finally, in cases $(e)$ and $(f)$, the amplitud is proportional to the Dirac mass. We stress that in the present model, the arguments in Ref.~\cite{Senjanovic:2015yea} 
which implies that the right-handed CKM matrix is close to the left-handed one, are not valid. Hence, not only the entries of  $V^L_{PMNS}$
but also those of $V^R_{PMNS}$ are free parameters. Notice that if neutrinos are Dirac particles, and since $m_{W_R}\gg m_\nu$ always, process in Fig.~\ref{fig:fig2}(b) is rather suppressed.

On the other hand, from Eqs.~(\ref{mcb5}) and (\ref{mnb4}) in the limit of $v_R$ being larger VEVs, we have
\begin{equation}
M^2_{Z_1}\approx \frac{M^2_{W_1}}{\cos^2\theta}+\mathcal{O}\left(\frac{X^2}{v^2_R}\right),
\label{mnb6}
\end{equation}
where $X$ is a VEV diferent from $v_R$, and
\begin{equation}
\cos^2\theta=\frac{\epsilon^2+\delta^2_L}{\epsilon^2+\delta^2_L+\epsilon^2\delta^2_L}.
\label{theta}
\end{equation}
Notice that when $\epsilon=1$ ($g_L=g_R\equiv g$, $g_{BL}=g^\prime$) we obtain
\begin{equation}
\cos^2\theta=\frac{g^2+g^{\prime\,2}}{g^2+2g^{\prime\,2}},
\label{theta2}
\end{equation} 
which is the value of this angle in left-right symmetric models~\cite{Chavez:2019yal}.
It means that $\theta$ can be identified, at least numerically,  with $\theta_W$ of the SM at a given energy, say, the $Z$-pole. 

In the case of the SM we know that weak interactions are more suppressed by the mass of $W$'s than by the coupling constant $g$. 
In fact $  \alpha_{QED}/\alpha_g = s ^ 2_W$, so this ratio is on the order of 0.231 i.e., $e/g$ is on the order of 0.48. 

However, if we use the exact expressions in Eqs.~(\ref{mcb2}) and  (\ref{mnb2}) we obtain
\begin{equation}
\frac{M^2_{Z_1}}{M^2_{W_1}}=
\frac{(1+\epsilon^2)x+(\delta^2_L+\epsilon^2)-\sqrt{\Delta_N}}{(1+\epsilon^2)x+\epsilon^2-\sqrt{\Delta}},
\label{mnb7}
\end{equation}
where $\Delta$ and $\Delta_N$ were defined by Eqs.~(\ref{delta}) and (\ref{deltapp}), respectively. We will assume $Z_1\approx Z$ and $Z_2\approx Z_R$; and $W_1\sim W_L$, $W_2\sim W_R$,
with the experimental value $M_W/M_Z=0.88147\pm0.00013$~\cite{pdg}. 

We have considered three cases: i) when $g_L=g_R$ (or $\epsilon=1$): If $v_R>3$ TeV, then $\delta_L\gtrsim 0.62$, $M_{W_2}> 960$ GeV and $M_{Z_2}> 1340$ GeV; ii) when $g_R$ is slightly smaller than $g_L$, say $\epsilon=0.8$: If $v_R>3$ TeV, then $\delta_L\gtrsim 0.71$, $M_{W_2}> 770$ GeV and $M_{Z_2}> 1090$ GeV; finally, iii) if $g_R > g_L$ and with increasing $v_R>3$ TeV, $\delta_L\to 0.54$, $M_{W_2}> 9620$ GeV and $M_{Z_2}> 96510$ GeV, i.e. extremely large values. 
\begin{center}
\begin{figure}[!h]
\includegraphics[width=15cm]{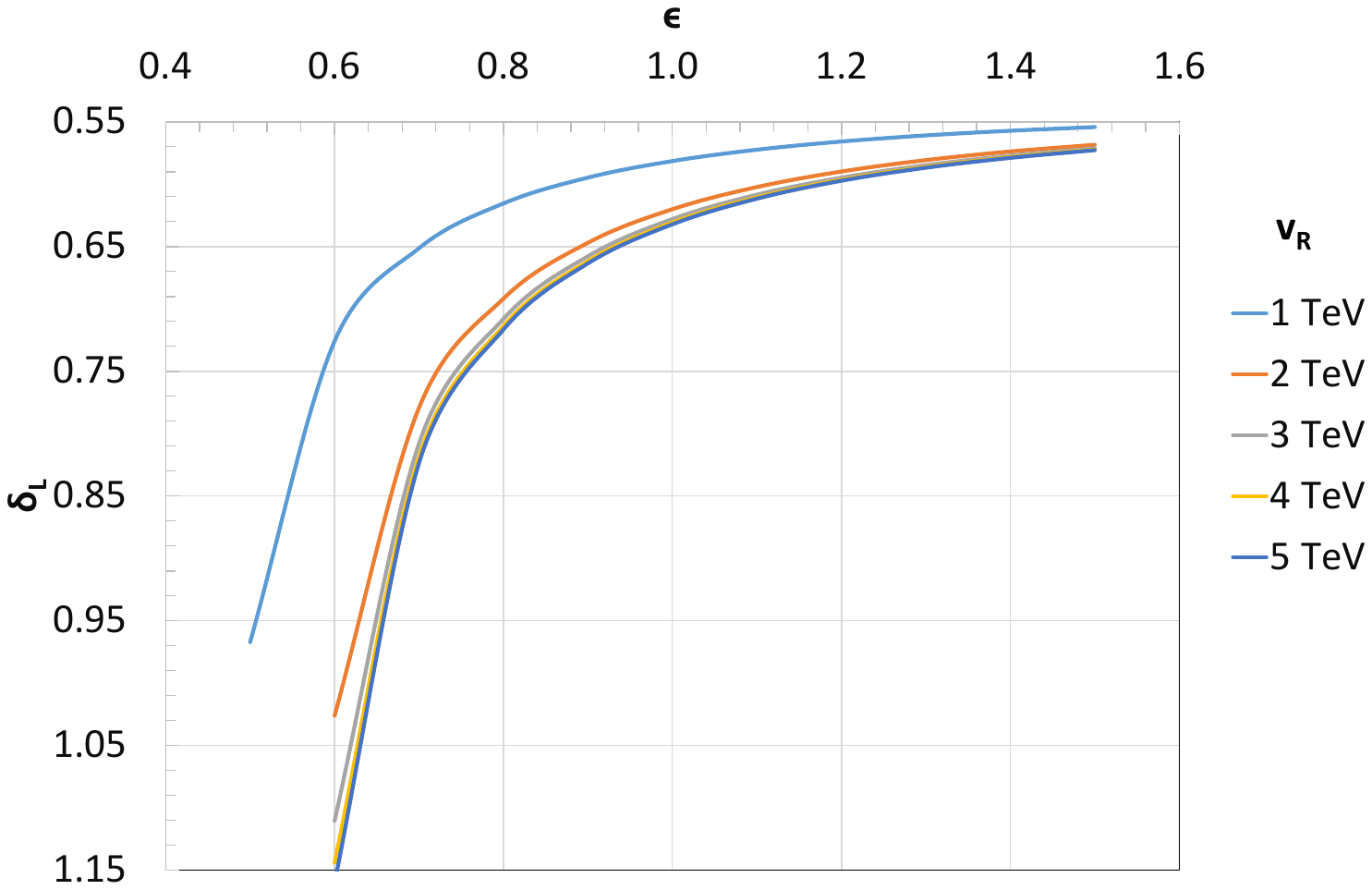}
\caption{Variation of $\delta_L$ with respect to $\epsilon$ for possible values of $\delta_L$ and $\epsilon$ in the range of W/Z mass ratio. We do not show the uncertainty bars because they are too small to be displayed.}
\label{fig:fig3}
\end{figure}
\end{center}

The graphic in Fig.~\ref{fig:fig3} shows the results for possible values of $\epsilon$ and $\delta_L$ obtained from Eq.~(\ref{mnb7}) and the $W/Z$ ratio from Particle Data Group, PDG data, with one standard deviation~\cite{pdg}. All those combinations of $\epsilon$ and $\delta_L$ showed in the graph make the mixing angle of this model approaches SM Weinberg angle, $\theta\rightarrow\theta_W$. We may conclude from this running that for increasing values of $v_R$ and $g_L=g_R$  ($ \epsilon=1$ ) then $g_{BL}  \approx (0.6 \pm 0.05)g_L$.
We can also say that for values of $g_R$ much greater than $g_L$, then $g_ {BL}$ tends to $0.5 g_L$.

If we were to introduce scalar triplets, neutrinos would be Majorana particles and the neutrinoless doubly beta decay $(\beta\beta)_{0\nu}$ could occur.
In this sort of model the decay has several contributions: active neutrinos, doubly charged scalars, or purely right-handed current via the exchange of right-handed neutrinos.
The inverse of the half-life $T^{-1}_{1/2}$ for this decay is proportional, not including the phase espace and the nuclear matrix element, to a dimensionless factor, say $\eta^2$. Let us consider just for illustration, the case of the right-handed current. In this case~\cite{Deppisch:2017vne}
\begin{equation}
\eta_{RHC}=m_p\left(\epsilon\,\sqrt{\zeta}\,\frac{M_{W_1}}{M_{W_2}}\right)^4\sum_i
\frac{(V^R_{PMNS})^2_{ei}M_i}{\vert p\vert^2+M^2_i},
\end{equation}
where $m_p$ is the proton mass, $\epsilon$ is the parameter defined in Eq.~(\ref{delta}), $V^R_{PMNS}$ is the right-handed mixing matrix, $\zeta^2$ is a dimensionless parameter that indicates how much the largest element of $(V^R_{PMNS})^2$ differs from $(V^L_{PMNS})^2$, and $M_i$ is the mass of the right-handed neutrino. Cases in which $M_i$ was larger or smaller than $p\approx 100$ MeV, using $\epsilon=1$, $M_{W_2}=5$ TeV, and $V^R_{PMNS}=V^L_{PMNS}$ were considered in Ref.~\cite{Deppisch:2017vne}. 
Here, we only want to see this particular case if $\epsilon\not=1$ and $V_R\not=V_L$. 
Note that if $\epsilon\sqrt{\zeta}=0.2$, the results of Ref.~\cite{Deppisch:2017vne} obtained with $M_{W_2}=5$ TeV can be obtained with $M_{W_2}=1$ TeV. 

Moreover, in the present model the left-handed triplets, $\Delta_L$, are not mandatory, and if included, they may be inert. Hence, only the type-I see-saws can be implemented through the interaction given in Eq.~(\ref{yukawa1}) plus the interaction of the complex scalar triplet $\Delta_R$ with leptons $\overline{(R^\prime)^c}\vec{\epsilon\tau}\cdot\Delta_RR^\prime$, (We omit flavour indices). If we do not introduce the triplet $\Delta_R$, neutrinos are then Dirac particles~\cite{Chavez:2019yal}.

Although it is not the objective of this article to make a detailed analysis of the phenomenology of this model, we  consider where to look for effects that differentiate it from other models of the same type. 
There are some tensions in the data of flavor physics. Just to mention an example, the determination of the elements $\vert V_{ub}\vert $ and $\vert V_{cb}\vert$ of the CKM matrix do not coincide when using the semi-lepton decays of the meson $B$. This discrepancy and some decays of the tau lepton may be indicators of a violation of lepton universality~\cite{Gambino:2020jvv}. In general, the LR symmetric model have the potential to solve these problems, and this is even more the case for the model considered here, in which there are more free parameters. However it is  interesting to search for constraints imposed by very well known decays, such as $\pi\to l\nu_L$
since, besides the contribution of the $W^-$ there are contributions of the physical singly charged scalars, say $H^-$. For instance, the interactions $Y^L\bar{l}_RG\nu_LH^-$ and $Y^R\bar{l}_LF\nu_RH^-$ induce effective interactions and constraints on $Y^{L,R}$ and $m_H$ can be obtained~\cite{Campbell:2008um,Crivellin:2013wna}. Recently, even considering only the active left-handed neutrinos, it was possible to find non-trivial regions for the Yukawa parameters and the mass of a singly charged scalar that are compatible with the measured ratio $\Gamma(\pi\to\mu \nu_\nu)/\Gamma(\pi\to l \nu_e)$~\cite{Guzzo:2021tpx}. In the present model there are also right-hand neutrinos and at least three charged scalars. Of course, these processes deserve a more detailed study and it will be shown elsewhere.

\section{Conclusions}
\label{sec:con}

Although we have considered Dirac neutrinos, if we were to add a triplet that transforms according to $\Delta_R\sim(\textbf{1}_L,\textbf{3}_R,+2)$ to replace the doublet $\chi_R$, the type I seesaw mechanisms could be implemented and neutrinos would be Majorana particles. In fact, since the nature (Dirac or Majorana) of neutrinos is not yet known, it would be interesting to compare this to in the context of the present model, for instance, in the production of $W^\pm_R$~\cite{Senjanovic:2016vxw,Senjanovic:2018xtu}. 

The most stringent limits on the $W_R$ mass was obtained in Ref.~\cite{Sirunyan:2018pom} which assumed that it decays into an electron (or a muon) and a heavy neutrino of the same lepton flavor in a strict left-right symmetry i.e., $g_R=g_L$ and also that the left- and 
right- mixing matrices in the lepton and quark sectors are equal. Under these conditions, at $\sqrt{s}=8$ TeV, $W_R$ masses of up to nearly 3 TeV were excluded. However, in the present model considered here, this mass may be smaller if $g_R\not=g_L$ and also the left- and right- mixing matrices are different from each other.

By considering $g_R< g_L$ and $V^L_{CKM}\not= V^R_{CKM}$ in Ref.~\cite{Frank:2018ifw} it was shown  
that the $W_R$ mass bounds can be considerably relaxed, while $Z_R$ mass bounds are much more stringent. This
reduces the $W_R$ production and decay cross-section which implies a smaller symmetry energy scale in the scalar spectrum
(triplets $\Delta_L$ and $\Delta_R$), which is not used in our case because there is no relation at all between the left- and right- scalars. 

This means that $W^\pm_R$ and $Z_R$ in the present model cannot be considered as sequencial $W^\prime,Z^\prime$, and the bounds on their masses 
obtained by ATLAS (A Toroidal LHC Apparatus)~\cite{Aaboud:2017yvp} and CMS (Compact Muon Solenoid)~\cite{Sirunyan:2018xlo,Sirunyan:2018pom} are not straightforwardly applicable to these
vector bosons. ThSese bounds strongly depend on the coupling of the vector bosons to quarks (and leptons) and in the model considered here, $g_R$
has no relation with $g_L$; the same is true of $V^L_{PMNS}$ and $V^R_{PMNS}$. Moreover, if neutrinos are Majorana particles (the triplet
$\Delta_R$ is introduced) the analysis depend on the assumption that $m_{\nu_R}$ is heavier or lighter than $W_R$. Note that,
in the present model plus a triplet, there is a doubly charged scalar $\Delta^{++}_R$ but not necessarily $\Delta^{++}_L$. The lower bounds on the mass of 
 $\Delta^{++}_L$ are between 383 GeV  and 408 GeV~\cite{Chatrchyan:2012ya}. Nevertheless, if parity is explicitly violated, $\Delta^{++}_L$ 
may be part of an inert triplet which may or may not be coupled with leptons. In this case, the type I seesaw mechanism can be implemented 
by the bi-doublet and the triplet $\Delta_R$; hence, the Yukawa couplings of the triplet $\Delta_L$, if allowed, are not proportional
to the neutrino mass matrix. 

Note that although our model, similarly to those in Refs.~\cite{Langacker:1989xa,Frank:2010cj}, allows 
$g_L\not= g_R$, $V^L_{CKM}\not= V^R_{CKM}$, and $V^L_{PMNS}\not= V^R_{PMNS}$,
it is still different from the model considered those references since we have a particle spectrum that is not even left- right- symmetric in the scalar multiplets and which also has no parity restoration.

\section*{ACKNOWLEDGMENTS} 

HD thanks to CONCYTEC for financial support  and to the IFT-UNESP for the kind hospitality provided for part of this work.
VP would like to thanks FAPESP under the funding Grant No. 2014/19164-6 and, the Faculty of Sciences of the Universidad Nacional de Ingenierıía (UNI) for the kind hospitality. OPR would like to 
thank FAPESP grant 2016/01343-7 for funding our visit to ICTP-SAIFR from 01-08/12/2019 where part of this work was done,
and, not least, to IFT-UNESP for the kind hospitality. In addition, E. Castillo thanks to OGI-UNI for financial support and to the IFT-UNESP for the kind hospitality.

\end{document}